\documentclass[
    ,final            
  ]
  {aipproc}

\layoutstyle{6x9}

\usepackage{amssymb}
\usepackage{amsmath}

\newcommand{\ten}{{\bf 10}}
\newcommand{\bfive}{{\bf \bar 5}}
\newcommand{\five}{{\bf 5}}

\begin{document}

\title{Extracting Flavor from Quiver Gauge Theories}

\classification{12.10.Dm, 11.25.Mj, 12.60.Jv, 14.60.Pq}
\keywords      {GUT, supersymmetry, D-branes, quiver, neutrino,
   Froggatt Nielsen, anarchy}

\author{Tomer Volansky}{
  address={Department of Particle Physics, Weizmann Institute of Science, Rehovot 76100, Israel}
}

\begin{abstract}
  We consider a large class of models where an $SU(5)$ gauge symmetry
  and a Froggatt-Nielsen (FN) Abelian flavor symmetry arise from a
  quiver gauge theory.  Such quiver models are very
  restrictive and therefore have strong predictive power.  In particular, under mild
  assumptions neutrino mass anarchy is predicted.
\end{abstract}

\maketitle


\section{Introduction - The Neutrino Flavor Puzzle}
\label{sec:introduction}

The flavor puzzle in the SM, namely the smallness and hierarchical
structure of the charged fermions, hints towards new
physics beyond the SM.  In particular, these features are very
suggestive that an approximate horizontal symmetry is at work.  A
solution that employs such a symmetry is the Froggatt-Nielsen (FN)
mechanism \cite{Froggatt:1978nt}. The various generations carry
different charges under a spontaneously broken Abelian symmetry. The
breaking is communicated to the SM fermions via heavy fermions in
vector-like representations and the ratio between the scale of
spontaneous symmetry breaking and the mass scale of the vector-like
fermions provides a small symmetry-breaking parameter. Yukawa
couplings that break the FN symmetry are suppressed by powers of the
breaking parameter, depending on their FN charge.

In recent years, the discovery of neutrino masses and the measurements
of several neutrino flavor parameters have introduced a new twist to
the flavor puzzle.  Indeed, the measured dimensionless parameters
(encoding the information on flavor physics) are all of order one,
implying a possible anarchy in the neutrino sector \cite{Hall:1999sn}.
The flavor puzzle must then also address the question: Why are the
neutrino flavor parameters different from those of the quarks and
charged leptons?

It is simple to write down a FN model which generates hierarchy in the
charged fermions and anarchy in the neutrino sector.  Consider for
example, an $SU(5)$ model with a $U(1)_{\rm FN}$ horizontal symmetry
under which the matter content is charged as $\bfive_i(0,0,0)$,
$\ten_i(2,1,0)$, $H_{u,d}(0,0)$ and the additional scalar, $S$,
responsible for spontaneously breaking the $U(1)_{\rm FN}$ has charge
$-1$.  It is straightforward to check that for $\langle
S\rangle/M=\epsilon\simeq 0.05$ the resulting mass matrices (leaving
out order one coefficients): {\footnotesize
\begin{eqnarray}
  \label{eq:1}
    M_u \sim \langle H_u\rangle
  \left(\begin{array}{ccc}
      \epsilon^4&\epsilon^3&\epsilon^2\\ \epsilon^3&\epsilon^2&\epsilon \\ 
      \epsilon^2&\epsilon&1  
    \end{array}
  \right),\qquad
  M_d \sim \langle H_d\rangle
  \left(\begin{array}{ccc}
       \epsilon^2&\epsilon^2&\epsilon^2 \\ \epsilon&\epsilon&\epsilon
  \\ 1&1&1 
    \end{array}
  \right),\qquad
  M_\nu \sim \frac{\langle H_u\rangle^2}{M} 
  \left(\begin{array}{ccc}
      1&1&1 \\ 1&1&1 \\  1&1&1 \\ 
    \end{array}
  \right),
\end{eqnarray}
} agree with the measured values.  Here and below, $M$ is the scale at
which the breaking is communicated to the SM.  On the other hand, it
is just as simple to generate a different structure in the neutrino
sector.  Taking a different set of charges, say $\ten_i(4,2,0)$ and
$\bfive_i(1,1,0)$, one finds that a different mass structure arises
but which still agrees with the data for $\epsilon\simeq 0.23$ and
different order one coefficients.

The reason for this freedom stems from the nature of the FN mechanism.
Indeed, model building within the FN framework usually requires a
value for the small symmetry-breaking parameter(s), and a set of FN
charges for the fermion and Higgs fields. These choices determine the
parametric suppression of masses and mixing angles. One then checks
that the experimental data can be fitted with a reasonable choice of
order-one coefficients for the various Yukawa couplings.  Thus all FN
predictions are subject to inherent limitations: 1) The FN charges are
not dictated by the theory.  2) The value of the small parameter is
not predicted. 3) There is no information on the ${\cal O}(1)$
coefficients.  The predictive power of the FN framework is thus
limited.

To make further progress, one would like to embed the FN mechanism in
a framework where some or all of the inherent limitations described
above are lifted. This may happen in string theory.  Below we consider
string-inspired models in which the FN mechanism is embedded in quiver
gauge theories \cite{Antebi:2005hr}. These theories arise at low
energy as the effective theories on D-branes placed at singular
geometries \cite{Douglas:1996sw,Johnson:1996py}. The structure of these
theories tightly constrains model building and hence the realization
of the FN mechanism.

\section{Embedding FN in Quiver Gauge Theories}
\label{sec:quiv-gauge-theor}

A quiver diagram is an efficient way for describing the gauge theory
obtained from the open string sector.  For oriented strings, nodes in
the quiver denote $U(N)$ gauge factors and the fields are represented
by directed lines connecting two such nodes. A line coming out of a
node stands for a field in the
fundamental representation, while a line going into a node represents a
field in the antifundamental. A line starting
and ending on the same node, describes a field in the adjoint
representation of the corresponding $U(N)$ factor.

The quiver diagram can be generalized to accommodate unoriented
strings.  Nodes represent $U(N)$, $SO(N)$ or $Sp(N)$ gauge groups and
the lines are no longer directed but instead, each line must be drawn
with an arrow at each of the two ends indicating what is the
representation of the corresponding string under each of the two gauge
group factors.  Unoriented strings with both ends coming out of the
same set of branes may reside in either the symmetric or the
antisymmetric combination of ${\bf N}\times \bf{N}$. 

A crucial property of quiver gauge theories is that the $U(1)$ charge
at each $U(N)=SU(N)\times U(1)$ gauge factor depends on the
representation under the non-Abelian part of the group.  In particular
a fundamental of the $SU(N)$ is charged $+1$ under the corresponding
$U(1)$ while an anti-fundamental is charged $-1$.  This property
strongly restricts the possible Abelian charges in a particular theory
and therefore plays a crucial part in the embedding of the FN
mechanism.  Without going to any details we note that many of these
$U(1)$ factors are in fact global where the corresponding gauge fields
obtain a mass through couplings to axions.  Such a higgsing process
occurs both for anomalous U(1)s through the generalized Green-Schwartz
mechanism \cite{Douglas:1996sw,Green:1984sg,Dine:1987xk}, and
upon compactification also to many non-anomalous U(1) factors
\cite{Buican:2006sn}. 

With the aid of the global $U(1)$s, embedding the FN is an easy
matter.  Here we restrict ourselves to a single $U(1)_{\rm FN}$
and therefore by assumption there is only one field in the quiver,
$S$, which spontaneously breaks the symmetry and obtains a {\it small} VEV,
$\langle S\rangle/M = \epsilon \ll 1$.  
As shown in \cite{Antebi:2005hr}, under these assumptions and due to
the relation between the abelian and non-abelian charges, the largest
possible suppression in such FN models is $\epsilon^3$.  This result
shows the strong predictive power that is added to the FN mechanism
when embedded in string theory.

In fact these constraints are so strong that they typically pose
phenomenological problems, essentially since the global $U(1)$
symmetries of the SM or its GUT extensions are not directly related to
the local gauge groups.  Consider as an example, an $SU(5)$ theory
with the following interactions:
\begin{equation}
  \label{eq:17}
  W = Y_{ij}^d\;H_d\cdot\ten_i\cdot\bfive_j + Y_{ij}^u\;H_u\cdot\ten_i\cdot\ten_j + (Y_{ij}^\nu/M)\;
  H_u\cdot H_u\cdot\ten_i\cdot\ten_j. 
\end{equation}
\begin{figure}[bt]
  \centering
  \includegraphics[scale=0.4]{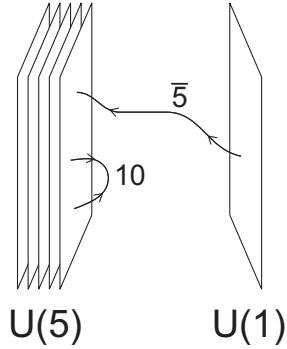}
  \caption{A D-brane construction with an $SU(5)$ gauge group and a
    distinct $U(1)_{\rm FN}$. The $\bfive$-plets are strings
    stretching between
    the two stacks and are thus charged under the FN group, while the
    $\ten$-plets  connect only to the $U(5)$ stack and have no
    $U(1)_{\rm FN}$ charge. }
  \label{fig:branes}
\end{figure}
As shown in figure \ref{fig:branes} the $\ten$-plets are generated by
strings having both ends on the $U(5)$ stack of branes.  Thus these
$\ten$-plets have charge $+2$ under the corresponding
$U(1)$.  Similarly, $H_u$ is a fundamental of $SU(5)$ and therefore
has a +1 charge under the $U(1)$.  With these charges no up-type
masses are allowed.  These problems are very generic in D-brane
constructions.  

There are three possible solutions to the above problem:
\begin{enumerate}
\item The particle content is extended in
  such a way that the symmetry is realized.  For example, an extended
  higgs sector \cite{Ibanez:2001nd} may be introduced.
\item The $U(1)$ is broken spontaneously \cite{Antebi:2005hr}.  The only way to do this
  without breaking the associated $SU(N)$ gauge group is by
  considering a composite singlet composed of $N$ fundamentals to obtain a VEV.
\item The $U(1)$ is broken by non perturbative effects.  
  Depending on the matter content, gauge theory instantons may be
  generated or otherwise stringy
  effects such as D-instantons may break the U(1)
  \cite{Antebi:2005hr,Buican:2006sn}.
\end{enumerate}
These non-perturbative effects relax some of the constraints on the
possible suppressions stated above and in particular allow an
$\epsilon^4$-suppression of the Yukawa couplings in the case of
$SU(5)$ GUT theory.

\section{$SU(5)$ and Anarchy}
\label{sec:su5-anarchy}
The problem discussed above also demonstrates that for the simple
$SU(5)$ theory all $\ten$-plets have equal charges under the FN
symmetry.  Thus, this scenario gives rise to up mass anarchy and is
therefore phenomenologically excluded.  One therefore must consider
product groups whereby an $SU(5)$ GUT model we mean that there is a
range of energy scales where the gauge group is $SU(5)$, with matter
fields that transform as ${\bf5}$, ${\bf\bar 5}$, and ${\bf10}$.
Product groups are a generic prediction of such FN models
\cite{Antebi:2005hr}.

The simplest models, on which we focus, have the following pattern of
gauge symmetry breaking: $SU(5)\times SU(5)\rightarrow SU(5)_{\rm
  diag}$.  The breaking is carried out by the FN field which is
charged $(+1,-1)$ under the corresponding $U(1)_{\rm L}\times
U(1)_{\rm R}$.  More complicated breaking patterns have a similar
hierarchical form, but involve extended particle content.
The $\bfive$-plets then transform under the $SU(5)\times
SU(5)\times U(1)_{\rm L}\times U(1)_{\rm R}$ as either
$(\bfive,1)_{-1,0}$ or $(1,\bfive)_{0,-1}$. The $\ten$-plets transform
as either $(\ten,1)_{+2,0}$ or $(1,\ten)_{0,+2}$ or
$(\five,\five)_{+1,+1}$.  The $H_u(\bf5)$ field transforms as
either $(\five,1)_{+1,0}$ or $(1,\five)_{0,+1}$.

The strongest mass hierarchy in the various fermion mass matrices
appears in the up sector. A viable model must produce this mass
hierarchy and it is straightforward to check that such hierarchy
requires the use of all three possible charges for the $\ten$-plets.
One is therefore left with no freedom in the choice of charges for
these fields.

For the down sector, since the $\bfive$ fields carry charges of either
$(-1,0)$ or $(0,-1)$, at least two of them have the same FN charge.
Thus, there must be at least quasi-anarchy in the
neutrino sector.  Such a situation has implications for the down
sector: either one or all three
 down mass ratios are of the same order
as the corresponding mixing angles ({\it e.g.} $m_s/m_b\sim|V_{cb}|$).
If one further requires the correct hierarchy in the down sector, one
finds that all $\bfive$-plets must have the same FN charge.  Thus 
  anarchy is obtained in the neutrino sector.  We note that this
prediction is independent of the choice for strong dynamics which
generate up-type masses.  From this point of view there is only one
viable quiver which is unique. The quiver is shown in \cite{Antebi:2005hr}.

\begin{theacknowledgments}
  I thank my co-authors Yaron Antebi and Yossi Nir for critical
  reading of this note.  This work is supported by a grant from the
  United States-Israel Binational Science Foundation (BSF), Jerusalem,
  Israel.
\end{theacknowledgments}






\end{document}